%% file: main.tex
\documentclass[twoside]{article}
\usepackage{slashed, graphicx, amssymb, amsmath, ragged2e, multicol, multirow, float, tabu, cite}

\usepackage{blindtext} 
\usepackage{url, bm, subfloat, makecell, subfigure}
\usepackage[sc]{mathpazo} 
\usepackage[T1]{fontenc} 
\usepackage{microtype} 

\usepackage[english]{babel} 

\usepackage[hmarginratio=1:1,top=32mm,columnsep=20pt,margin=2.3cm]{geometry} 
\usepackage[hang, small,labelfont=bf,up,textfont=it]{caption} 
\usepackage{booktabs} 

\usepackage{lettrine} 

\usepackage{enumitem} 
\setlist[itemize]{noitemsep} 

\usepackage{abstract} 

\usepackage{titlesec} 
\renewcommand\thesection{\Roman{section}} 
\renewcommand\thesubsection{\roman{subsection}} 
\titleformat{\section}[block]{\large\scshape\centering}{\thesection.}{1em}{} 
\titleformat{\subsection}[block]{\large}{\thesubsection.}{1em}{} 

\usepackage{fancyhdr} 
\usepackage{titling} 
\graphicspath{{chap/chapter1/fig/}{fig/}}

\usepackage{subfiles}

\renewenvironment{abstract}
 {\small
  \begin{center}
  \bfseries \abstractname\vspace{-.5em}\vspace{0pt}
  \end{center}
  \list{}{
    \setlength{\leftmargin}{.5cm}%
    \setlength{\rightmargin}{\leftmargin}%
  }%
  \item\relax}
 {\endlist}

\pretitle{\begin{center}\huge\bfseries} 
\posttitle{\end{center}} 
\title{Failure modes and downtime of radiotherapy linear accelerators and Multi-Leaf Collimators in Indonesia} 
\author{%
\textsc{G. Peiris$^{1}$, S. Pawiro$^{2}$, M. Kasim$^{2}$, S. L. Sheehy$^{3}$} \\[1ex] 
\normalsize \emph{$^{1}$School of Physics, University of Melbourne, Australia} \\ 
\normalsize \emph{$^{2}$Faculty of Mathematics and Natural Sciences, University of Indonesia, Indonesia} \\ 
\normalsize \emph{$^{3}$Department of Physics, University of Oxford, England} \\ 
}

\date{} 
\renewcommand{\maketitlehookd}{%
\begin{abstract}
\noindent 
The lack of equitable access to radiotherapy linear accelerators (LINACs) is a substantial barrier to cancer care in Low and Middle-Income Countries (LMICs). Aside from the issue of cost, there are also issues of robustness of state-of-the-art LINACs in LMICs, which are subject to mechanical and electrical breakdowns, which can result in downtimes ranging from days to months. Recent research has identified disparities in failure frequency and downtimes between LMICs (Nigeria, Botswana) and a High-Income Countries (HIC, the UK), and highlighted the need for additional data and study particularly relating to Multi-Leaf Collimators (MLCs). This study analyses data from 14 Indonesian Hospitals with a total of 19 LINACs and shows the pathways to failure of radiotherapy LINACs and frequency of breakdowns with an additional focus on the Multi-Leaf Collimator (MLC) subsystem. We found that LINACs throughout Indonesia are out of operation for 7 times longer than HICs and the Mean Time Between Failures of a LINAC in Indonesia is 341.58 Hours, or about 14 days. Furthermore, of the LINACs with an MLC fitted, 5$59.02^{+1.98}_{-1.61}\;\%$ of mechanical faults are due to the MLC and $57.14_{-1.27}^{+0.78}\;\%$ of cases requiring a replacement component are related to the MLC. These results outline the need to reassess the current generation of RT LINACs and ultimately work towards guiding future designs to be robust enough for all environments. 
\\

\emph{Keywords}: Radiotherapy, Multi-Leaf Collimator, Downtime
\end{abstract}
}

\begin{document}
\maketitle

\section{Introduction}  \label{Introduction}
\subfile{sections/Introduction/introduction}

\section{Material and Methods}\label{Materialandmethods}

\subfile{sections/Materialandmethods/materialandmethods}

\section{Results}\label{Results}

\subfile{sections/Results/results}

\section{Discussion} \label{Discussion}
\subfile{sections/Discussion/discussion}

\section{Conclusion} \label{Conclusion}
\noindent
This study investigated the state of radiotherapy LINACs in Indonesia with a focus on the MLC subsystem and shows $59.02^{+1.98}_{-1.61}\;\%$ of mechanical faults and $57.14_{-1.27}^{+0.78}\;\%$ of replacements are due to the MLC. A LINAC in Indonesia is down for 7 times longer than one in the UK overall and an MLC is down for 4.27 times longer. Though the MTBF of a LINAC is 341.58 hours, the median is just 52.5 hours. This work supports and quantifies earlier research findings that LMICs as especially susceptible to mechanical faults and prolonged downtimes in radiotherapy LINACs. 
Determining suitable alternatives or updates to Multi-Leaf Collimator designs requires a representative sample of the state of radiotherapy in Low- and Middle-Income Countries and these results provide an attempt to produce this data. Recommendations for a more comprehensive method for record keeping and potentially reducing the number of leaves have been motivated. Providing better cancer care must entail a reassessment of the MLC to account for discrepancies in LINAC downtimes and failure modes between LMICs and HICs.

\newpage
\bibliography{LitRev.bib}{}
\bibliographystyle{ieeetr}


\end{document}


\section{Appendix A}
\subsection{The Stand}
	\subsubsection{Microwave Power Sources}
Within the LINAC stand, there needs to be a suitable method of generating power, specifically radiofrequency (RF) power to the accelerator. In high-energy LINACs, a klystron is used for power while a magnetron is used for low-power LINACs. Both the klystron and magnetron utilise microwave cavities in their operation. \\
Microwave cavities are cylindrical resonators with oppositely charged end caps, generating an E field through the $z$-axis of the cavity. An electron passing through the cavity will be accelerated by this E-field either by a travelling wave or standing wave. This E field generates a circular B field. This creates a natural focusing due to the electromagnetic force pushing an electron beam to the centre of the cavity. This principle is prevalent in klystrons and magnetrons as well as the accelerating structure.
		\paragraph{Klystron}
The klystron has multiple microwave cavities connected via drift tubes, a cathode on one end to provide electrons and a collector on the opposite end. An input microwave accelerates electrons from the cathode through the cavities to the collector. The last cavity, called the catcher, outputs an amplified high-power signal.\\
Because the input signal oscillates between positive and negative voltage, the electrons bunch up by the cathode in a positive phase and are accelerated as a bunch in the negative phase. It is important to note that the output RF Power doesn't increase linearly with the RF power put in. For a given cathode voltage, the klystron has a characteristic curve which determines the RF input power to maximise the output. A suitable mathematical model~\cite{JACMP} for a klystron, with power $P_{RF}$ input, is given by:
	\begin{align}
		P_{kly} = 10P_{Max}\left(\frac{J_1(X)}{X}\right)^2\frac{P_{RF}}{P_C}; && X = 1.84\sqrt{\frac{P_{RF}}{P_C}},
	\end{align}
	where $P_{Max}$ is the beam potential-dependent maximum output power, $P_C$ is the critical power level at which saturation occurs. The plot showing the RF In against RF Out for various cathode voltages is shown in Figure \ref{fig:klyfig}. The plot suggests that increasing the klystron voltage and having a lower RF input will yield a larger amplification or gain. In the case of the 125kV cathode, a 72W input gets amplified to 5.6MW, which is a gain of almost 49dB.\\
	\begin{figure}[h]
    	\centering
    	\includegraphics[scale=0.55]{KlyCharCurv.png}
		\label{fig:klyfig}
		\caption{Power output from the klystron vs. RF power driving it, for varying levels of cathode voltage.}
	\end{figure}\\ \\
	A klystron is very effective at amplifying an input signal with very little input voltage since accelerating small particles to high velocities doesn't take much power.
	
	\paragraph{Magnetron}
 Like the klystron, the magnetron has a cathode ejecting electrons towards an anode through an evacuated drift space. One key difference is the circular geometry of the magnetron in contrast to the linear geometry of the klystron. Electrons from the hot filament are attracted to the outermost ring of the magnetron. However, the magnetic field orthogonal to the cross-section, in addition to the electric fields, diverts the electrons from their preferred path. This results in a rotating electron cloud which induces a resonant RF field in the cavity through the oscillation of charges. A short antenna attached to one of the spokes can extract the RF field.
 	\begin{figure}[h]
    	\centering
    	\includegraphics[scale=0.55]{magnetron.jpg}
		\caption{Cross section of a magnetron~\cite{eeeguide.com}.}
	\end{figure}
	\subsection{Waveguide}
	Waveguides convey the power generated by the klystron or magnetron to the accelerating structure within the gantry. In typical systems where power needs to be generated and conveyed elsewhere, the preference is to use coaxial cables. They are more flexible, can carry multiple signals and have a wide frequency range. Waveguides by comparison can be more expensive and cause time delays in construction and reconfiguration~\cite{Coax}. Despite this, waveguides have low losses and are able to handle a much higher power range. Waveguides are also much more mechanically durable in comparison to the oft delicate coaxial cables.
	 
	\subsection{Gantry}
	The rotating gantry section contains all of the accelerator components, with power being supplied by the relevant components in the stand. It is important that the gantry can rotate, so that healthy cells are hit with far less dosage than the cancerous cells, which receives radiations from all angles.
	\subsubsection{Accelerator Structures}
	A series of microwave cavities are strung together to accelerate a beam electrons to a specified energy. The component tasked with this is the linear accelerating structure, also called accelerator waveguide. The electrons emitted from the cathode are bunched in a manner similar to the klystron. Typically, however, only about one third of the injected electrons are captured and accelerated. Just as in the klystron, electron bunches are accelerated using an electric field. The strength of the field will vary depending on the length of accelerating structure and the desired output energy. For a LINAC length, $x$, and electron bunch energy, $U$, the electric field, $\vec{E}$, needed in the LINAC is given by:
	\begin{align}
		\vec{E} = \frac{U^2-m_e^2c^4}{2qxm_ec^2}
	\end{align}
	Typically, low energy LINACs $(U \sim 4$MeV) have a length of about 30cm, which higher energy units can be up to a meter long. In Figure \ref{fig:Efig}, we see the required electric field as a function of LINAC length for differing bunch energies. It is important to note that the equation above uses a simplified model and does not take into account space charge effects of the bunch, nor the images charges produced by the LINAC walls.\\
	\begin{figure}[h]
    	\centering
    	\includegraphics[scale=0.55]{E-Field.png}
		\caption{Electric field as a function of linac length.}
		\label{fig:Efig}
	\end{figure}
	Accelerator structures are often found in two varieties: Travelling Wave and Standing Wave. Both refer to the manner is which the electric field propagates through the series of cavities.
	
	\paragraph{Standing-wave Accelerating Structure}
	This machine links its microwave cavities such that each cavity alternates the direction of its electric field. Over time, each cavity oscillates between some $+E$ and $-E$ producing a "standing wave". As such, only every second cavity will have an electron beam. These accelerators tend to be more compact, requiring less space. However this means it has a high accelerating gradient and needs a stronger power source. It is for this reason standing wave accelerators are powered by klystrons.
		
	\paragraph{Travelling-wave Accelerating Structure}
	Unlike the standing wave, the travelling wave accelerator is longer with a shallower accelerating gradient. The electric field propagates linearly through the structure carrying bunches of electrons with it. A lower accelerating gradient means a magnetron is enough to power this machine.
	
	\subsection{Bending Magnet}
	Within the far end of the rotating gantry is an electromagnet which redirects the electron beam down towards the patient. It will either hit a tungsten target to convert the electron beam into an X-Ray distribution or continue into the patient, as per electron therapy. The bending magnet loops the beam through a 270$^{o}$ deflection. This is because such a bend is achromatic- regardless of the incoming energy of the beam, the output position and beam size will be the same~\cite{Karzmark}. This is in contrast to a 90$^{o}$ deflection, where variations in energy result in a spray of the electron beam. The distinction is visually apparent in the figures below.

\begin{subfigures}
		\begin{figure}[h]
  			\centering
    		\includegraphics[scale=0.5]{270bendingmagnet.png}
			\caption{270$^{o}$ bending magnet ensures the bean stays focused~\cite{Thwaites}.}
			\label{fig:270mag}
		\end{figure}
		\begin{figure}[h]
  			\centering
    		\includegraphics[scale=0.6]{90bendingmagnet.jpg}
			\caption{The 90$^{o}$ bending magnet will defocus the beam for particles with differing energies.}
			\label{fig:270mag}
		\end{figure}
	\end{subfigures}
	
	\subsection{Treatment Head}
	The purpose of the treatment head is to collimate and shape the X-Ray beam so that it will deliver the correct dosage of radiation to the appropriate areas- avoiding as much healthy tissue as possible. When the prescribed dosage is given, the dual ionisation chamber will produce an electrical to stop the treatment. Several collimators are used to block out peripheral X-Rays. One such collimator is the multi-leaf collimator (MLC).
	
	\paragraph{Multi-Leaf Collimator} The MLC is a set of 2 jaws each comprised of a series of thin leaves which move in and out. They conform to the shape of the tumour at a birds eye view with some able to change shape real time as the gantry rotates. A highly dense tungsten alloy is used in the leaves to prevent unwanted X-Rays from passing through the collimator.

%% file: sections/Introduction/introduction.tex
\begin{figure}[h]
		\centering
		\includegraphics[scale=0.4]{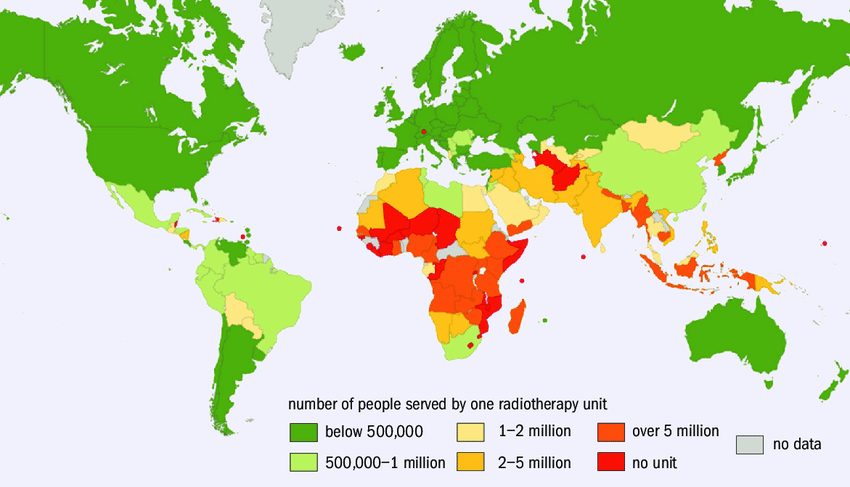}
		\caption{Many LMICs have limited access to radiotherapy units, displayed here. In some cases a single breakdown could mean an entire nation has no working radiotherapy machines until it is repaired.~\cite{Pistenmaa}}
		\label{fig:Pistenmaa}
	\end{figure}

Radiation Therapy is an effective and ubiquitous form of treatment used in modern cancer care. Where radiotherapy is available it is used in 40\% of all successfully treated cancer cases~\cite{RadOnc1}. State of the art radiotherapy relies on using a compact linear accelerator - or LINAC - with a typical operating lifetime of around 10 years in HIC clinics. There is a dramatic shortfall of LINACs in LMICs: it has been estimated that at least 5000 new LINACs in the next two or three decades are needed to meet the increasing burden of cancer in these regions~\cite{Pistenmaa2,Atun,Pistenmaa}. However, current technology is not well-suited to meet this need, with recent studies into the quality of cancer care in sub-Saharan Africa showing LINACs "often do not function well in the adverse conditions encountered in LMICs"~\cite{Pistenmaa2}. This observation has been substantiated in further research, which found that LINAC breakdowns in LMICs are more frequent and on average much longer than in HICs, with downtimes ranging from days to months~\cite{Reichenvater, Wroe}. 
Regions of low GDP are particularly vulnerable as they do not have sufficient LINACs to handle their patient load, as seen in Figure \ref{fig:Pistenmaa}. In cases where an LMIC has only one LINAC, its breakdown can be highly disruptive with patients either missing out on treatment or having to travel large distances to access the next available facility, which could be in another country. In addition to the issue of robustness, the upfront cost and maintenance costs of LINACs can be a substantial impediment in LMICs when considering the relative expense compared to - for example - personnel costs. 
A possible alternative to the LINAC is to supply mega-voltage X-rays using a radioactive source, usually cobalt-60 or caesium-137. Although these are inherently reliable and have few moving parts, due to safety and nuclear security issues they are considered as a less suitable technology option compared to LINACs for LMICs~\cite{Pistenmaa2} and are thus not considered in this study.
	\\
Many of the faults appear to be linked to inconsistent power supply and insufficient preventative maintenance, along with inaccessible spare parts and inadequate access to expert training courses~\cite{Reichenvater}. One of the particularly troublesome components is the Multi-Leaf Collimator (MLC) which is also reported to break down frequently in High Income Countries~\cite{Wroe,Kron,Wojtasik}.
However, to date research has not sufficiently investigated the extent to which MLCs are problematic in the delivery of radiotherapy in LMICs, primarily due to the lack of machines fitted with MLCs in the previous study based in Sub-Saharan Africa~\cite{Wroe}. 
In this work we analyse the recorded breakdown data of 19 LINACs from 14 hospitals across Indonesia, focusing on the MLC. The advantage of having all the data from a single country is the control of variables such as GDP and climate; all breakdowns across the various hospitals can be used as a representative sample of Indonesia as a whole.

%% file: sections/Materialandmethods/materialandmethods.tex
\begin{table*}[t]
\begin{center}
	\begin{tabular}{|| l  l  c  c  c ||} 
 		\hline
 		\textbf{Label} & \textbf{Hospital} & \textbf{No. of Linacs} & \textbf{Years of Data} & \textbf{No. of Patients} \\
 		\hline\hline
		A & RSUD Pasar Minggu & 1 & 2017 - 2019 & 10000 \\
		\hline
 		B & RS Pusat Pertamina Central Hospital & 1 & 2006 - 2019 & 45300 \\
 		\hline
 		C & RSUD Provinsi NTB & 1 & 2018 - 2019 & 3400 \\
 		\hline
 		D & RS Persahabatan Hospital & 1 & 2017 - 2019 & 16800 \\
 		\hline
		E & Dr Cipto Mangunkusumo General Hospital & 5 & 2000 - 2019 & 182400 \\
 		\hline
		F & RS Dr. Hasan Sadikin Central General Hospital & 1 & 2004 - 2018 & 208800 \\
 		\hline
		G & RSUD Abdul Wahab Sjahranie Hospital & 1 & 2016 - 2019 & 49400 \\
 		\hline
		H & RS Kanker Gading Pluit Hospital & 1 & 2013 - 2019 & 11520 \\
 		\hline
		J & RS General Hospital H. Adam Malik & 1 & 2013 - 2018 & 85800 \\
 		\hline
		K & Andalas University Hospital & 1 & 2019 & 4500 \\
 		\hline
		L & RS Murni Teguh Memorial Hospital & 2 & 2015 - 2019 & 58800 \\
 		\hline
		M & RSU Vina Estetica & 1 & 2014 - 2019 & 53400 \\
 		\hline
		N & MRCCC Siloam Hospital & 2 & 2014 - 2019 & 65000 \\
 		\hline
 		P & RSUD Indriati Hospital & 1  & 2017 - 2019 & 5700 \\ 
 		\hline
	\end{tabular}
	\caption{Summary of all Indonesian Hospitals used in the research and their number of LINACs. Number of Patients is an estimate extrapolated per facility across the years of data provided.}
	\label{table:indodata}
\end{center}
\end{table*}

\subsection{Collection of Indonesian hospital data} \label{dataColl}
The data used in this study was collected during a specialised workshop held in Jakarta, Indonesia in July 2019~\footnote{This workshop was generously supported by the University of Oxford Global Challenges Research Fund (GCRF) and enabled by the Indonesian Ministry for Health.}. Representatives (mostly medical physicists) from 14 hospitals across Indonesia were invited to use their LINAC logbook data to populate a template spreadsheet provided by the authors with instructions in both English and Indonesian. A survey was also conducted to collect relevant contextual data including staffing levels, number of LINACs and typical operating hours, and issues around power, humidity, and external environmental issues. The facilities involved in the study and the number of RT LINACs at each are presented in Table \ref{table:indodata} with data spanning from as far back as 2000. Each facility provided one or more logbooks of the machine's failures and maintenance as recorded by medical physicists or engineers.\\ \\
The date and duration of each fault was recorded as `Downtime' and the duration between faults was recorded as `Time Between Failures'. Each downtime length is sorted into one of 3 categories:
	\begin{itemize}
		\item \textbf{A} $\leq$ 5 mins,
		\item 5 mins < \textbf{B} $\leq$ 59 mins,
		\item \textbf{C} > 59 mins.
	\end{itemize}
A LINAC's `uptime' or `runtime' is defined by the total hours of operation from commissioning to decommissioning. \\ \\
This categorisation has been used in previous downtime studies~\cite{Wroe,SangGyu} and is justified based on disruption to workflow and irradiation capability. Faults in the LINACs were further categorised by their cause for failure: {Mechanical, Electrical, Board, Cabling, External, Parameter Drift}; and the method by which the fault was fixed: {Reset, Replace, Repair, Calibrate}. \\ \\
The logbook data is used in conjunction with the survey data. This allows the failure data to be normalised accounting for the differences in the ranges of data. All the relevant data has been normalised by 1000 patients treated which is more representative of a machine's usage than the hours of uptime. It should be noted that facilities which have multiple LINACs provided data on the average number of patients treated per machine each day. \\ \\
The logbooks varied significantly in the details provided. Most provided the date and time of the fault, a brief description of the fault and how it was repaired and a duration of the downtime. The differences in record keeping habits between facilities made the automation of data collection and analysis challenging. As such all data was checked for quality, categorised and entered into an Excel spreadsheet, later utilised as a csv file for processing with Python. There is a large variety in the vendors and models and not all logbooks begin at the commissioning of the machine.\\ \\
This particular set of data gives us a deeper insight into the behaviour of RT machines in LMICs with information such as the type of fault and the method used to rectify the error. In order to analyse the large dataset (i.e., 4900 faults recorded across 19 Indonesian LINACs over 19 years), two methods paralleling Wroe's comparative study~\cite{Wroe} were implemented. The first categorises and compares downtimes of LINACs and the time between failures and the second categorises the failure modes and methods used to repair LINACs.\\ \\

\subsection{Multi-Leaf Collimator Subsystem}
As this study is focused on the MLC, instances of MLC related failures were isolated in order to investigate how frequently this subsystem failed and to analyse its overall impact on LINAC downtime. The downtime lengths, reason for failure and resolution method are sorted using the same categories as described in Section \ref{dataColl} Publicly available data pertaining to each type of LINAC was used to identify the number of MLC leaves and leaf width for each machine, recorded in Table \ref{table:indomlcdata}. 

\begin{table*}[t]
\begin{center}
	\begin{tabular}{|| l  l  c  c  c  c  c ||} 
 		\hline
 		\textbf{Label} & \textbf{Vendor and Model} & \textbf{Year} 	& \textbf{Leaves} & \textbf{Leaf Width [mm]} & \textbf{Downtime} & \textbf{MLC Downtime}\\
 		\hline\hline
		A & Varian Trilogy 			& 2017 & 120 	& 5				& 23\%   	&  1.2\% \\
		\hline
		B & Siemens Primus$^\dagger$& 2006 &   		& 				& 11\%   	&		 \\ 
		\hline
		C & Varian Clinac CX 		& 2017 &  80 	& 5 			& 0.68\% 	& 13.4\% \\ 
		\hline
		D & Elekta Precise 			& 2016 &  80 	& 10 			& 4.9\%  	& 28.4\% \\ 
		\hline
 		\multirow{5}{*}{E} 	& Varian Clinac 2100C* & 2000 &  80 & 5 & 24\% 		&  5.7\% \\
 		\cline{2-7}
		  & Varian Clinac 2100C* 	& 2001 &  80 	& 5 			& 30\%		&  9.8\% \\
		\cline{2-7}
		  & Elekta Synergy S 		& 2006 &  80 	& 10 			& 17\%		& 12.2\% \\
		\cline{2-7}
		  & Elekta Synergy Platform & 2008 &  80 	& 10 			& 13\%	    & 83.9\% \\
		\cline{2-7}
		  & Varian Unique			& 2017 & 120 	& 5 			& 4.5\%     & 49.8\% \\
		\hline
		F & Elekta Precise$^\dagger$& 2004 &   		& 				&			&		 \\
		\hline
		G & Elekta Precise 			& -	   &  80 	& 10 			& 6.4\% 	& 19.5\% \\
		\hline
		H & Elekta Synergy Platform & 2012 &  80 	& 10 			& 6.5\% 	& 22.0\% \\
		\hline
		J & Elekta Precise 			& -    &  80 	& 10 			& 14\%  	& 26.8\% \\
		\hline
		K & Varian Clinac CX 		& 2016 &  80 	& 5 			& 9.0\% 	& 83.9\% \\
		\hline
 		\multirow{2}{*}{L} & Elekta Synergy Platform & 2013 &  80 & 10 & 0.95\% & 0.41\% \\
 		\cline{2-7}
		 & Elekta Versa HD 			& 2017 & 160 	& 5 			& 19\%  	& 0.02\% \\
		\hline
		M & Siemens Primus M 		& 2011 &  58 	& 10	 		& 14\%  	&  1.8\% \\
		\hline 
		N & Varian Clinac iX 		& 2010 &  80 	& 5 			& 2.6\% 	& 27.6\% \\
		\hline
 		P & Varian Clinac iX 		& 2017 &  80 	& 5 			& 1.5\% 	& 38.6\% \\ 
 		\hline
	\end{tabular}
	\caption{Types of LINACs commissioned at each facility and associated MLC parameters. All MLCs have a field size of 40$\times$40. Leaf widths denote the width of the central 20cm of field. * indicates machines which have been decommissioned at the hospital. $^\dagger$ indicates machines without an MLC. Downtime is the percentage downtime of the whole LINAC. MLC Downtime is the downtime percentage contribution of the MLC to the total downtime.}
	\label{table:indomlcdata}
\end{center}
\end{table*}

%% file: sections/Results/results.tex
\subsection{Analysis of Downtime in Indonesia}
The overview of LINACs in the study including percentage downtime caused by the MLC, number of MLC leaves and leaf width is shown in Table ~\ref{table:indomlcdata}. In Table~\ref{table:faultdata}, the LINAC faults include two machines (B, F) which do not have an MLC fitted. Though type C faults only make up 43\% of all faults by number, they contribute to over 98\% of the total downtime of all LINACs. While this implies that type C faults are more severe, it is caused in part by a lack of data on shorter interruptions as LMIC centres do not always record instances of type A and B faults. \\ \\
When only considering machines with an MLC fitted, 27.3\% of all Type C faults are related to the MLC. The average downtime experienced by a LINAC from a type C fault is 2.6$\pm$0.81 days. For the MLC subsystem, the average downtime is shorter, at 1.03$\pm$0.23 days. \\ \\
Wroe's investigation of LINACs in the UK found that the mean downtime for a disabling fault is 338.8 minutes. In stark comparison, LINACs in Indonesia have a mean downtime of 2391.76 minutes - 7 times longer than in the UK. 

\begin{table*}[t]
\begin{center}
	\begin{tabular}{|| l | l  l  l  l  l ||} 
 		\hline
 		 & \textbf{Category} & \textbf{Faults} & \textbf{\thead{Total \\ Downtime [hrs]}} & \textbf{\thead{Mean \\Downtime [mins]}} & \textbf{\thead{Median \\Downtime [mins]}} \\
 		\hline\hline
		\multirow{3}{*}{LINAC} 	& A & 102 	& 1.51 (0.002\%) 	  & 1.65    $\pm$ 0.162 	& 1 \\ 
 								& B & 2088 	& 798.61 (1.19\%) 	  & 22.95   $\pm$ 0.254	 	& 20 \\
 								& C & 1657 	& 66052.43 (98.80\%)  & 2391.76 $\pm$ 522.70	& 165 \\
		\hline
 		\multirow{3}{*}{MLC} 	& A & 37  & 0.900 (0.01\%) 		& 1.46    $\pm$0.27 	 	& 1 \\ 
 								& B & 733 & 297.43 (3.39\%) 	& 24.35   $\pm$0.42 	 	& 23 \\
 								& C & 438 & 8463.20 (96.60\%) 	& 1159.34 $\pm$338.19 		& 120 \\
 		\hline
	\end{tabular}
	\caption{Comparison of mean and median failure downtimes by downtime category in all LINACS and in the MLC. \textbf{A} constitutes a downtime less than 5 minutes, \textbf{B} a downtime between 5 minutes and 1 hour, and \textbf{C} downtimes are longer than an hour.}
	\label{table:faultdata}
\end{center}
\end{table*}

\subsection{Overview of Failure Rates and Reasons for Failure}
\begin{figure*}[h]
	\centering
	\begin{subfigure}
  		\centering
    	\includegraphics[scale=0.55]{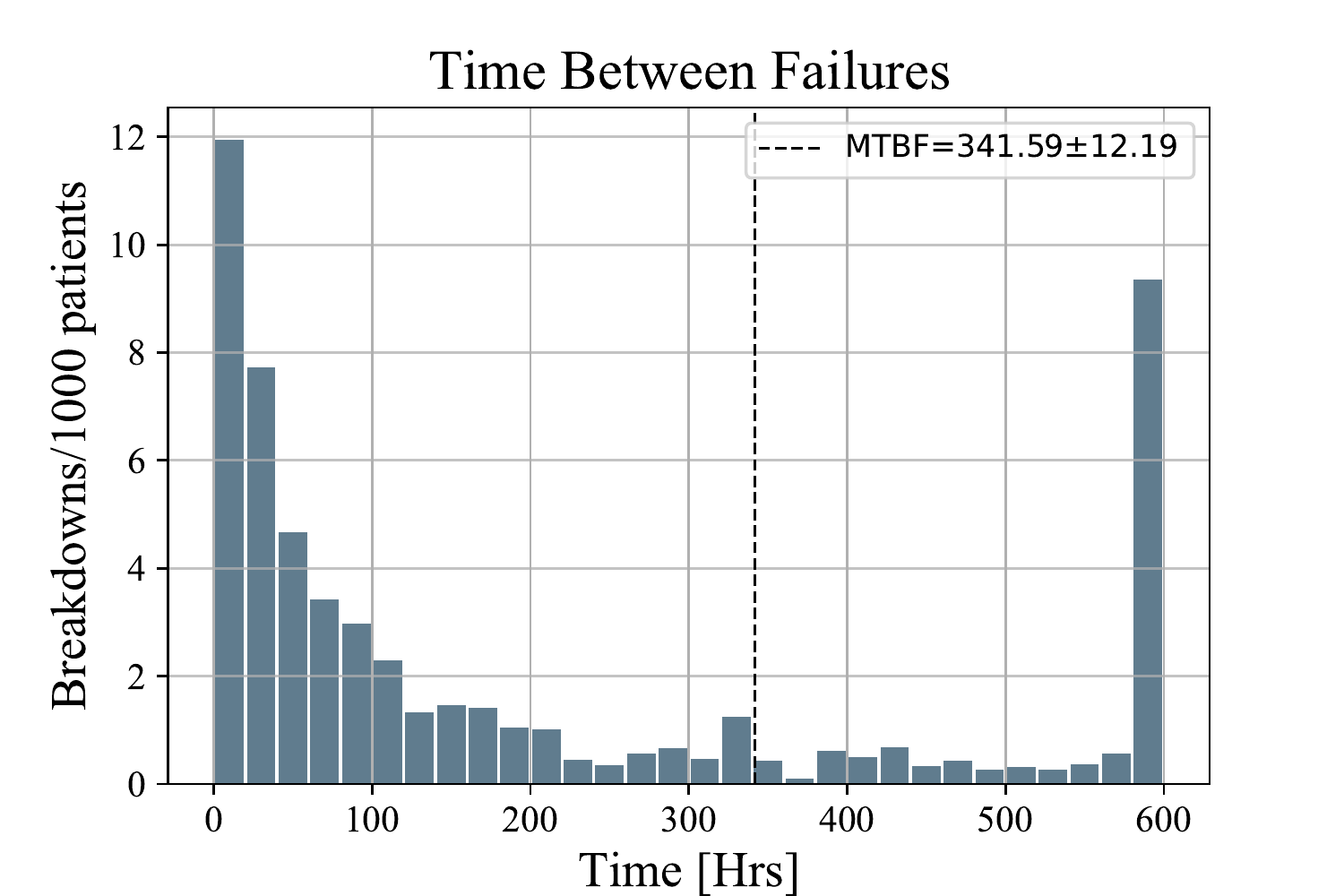}
	\end{subfigure}
	\begin{subfigure}
  		\centering
    	\includegraphics[scale=0.55]{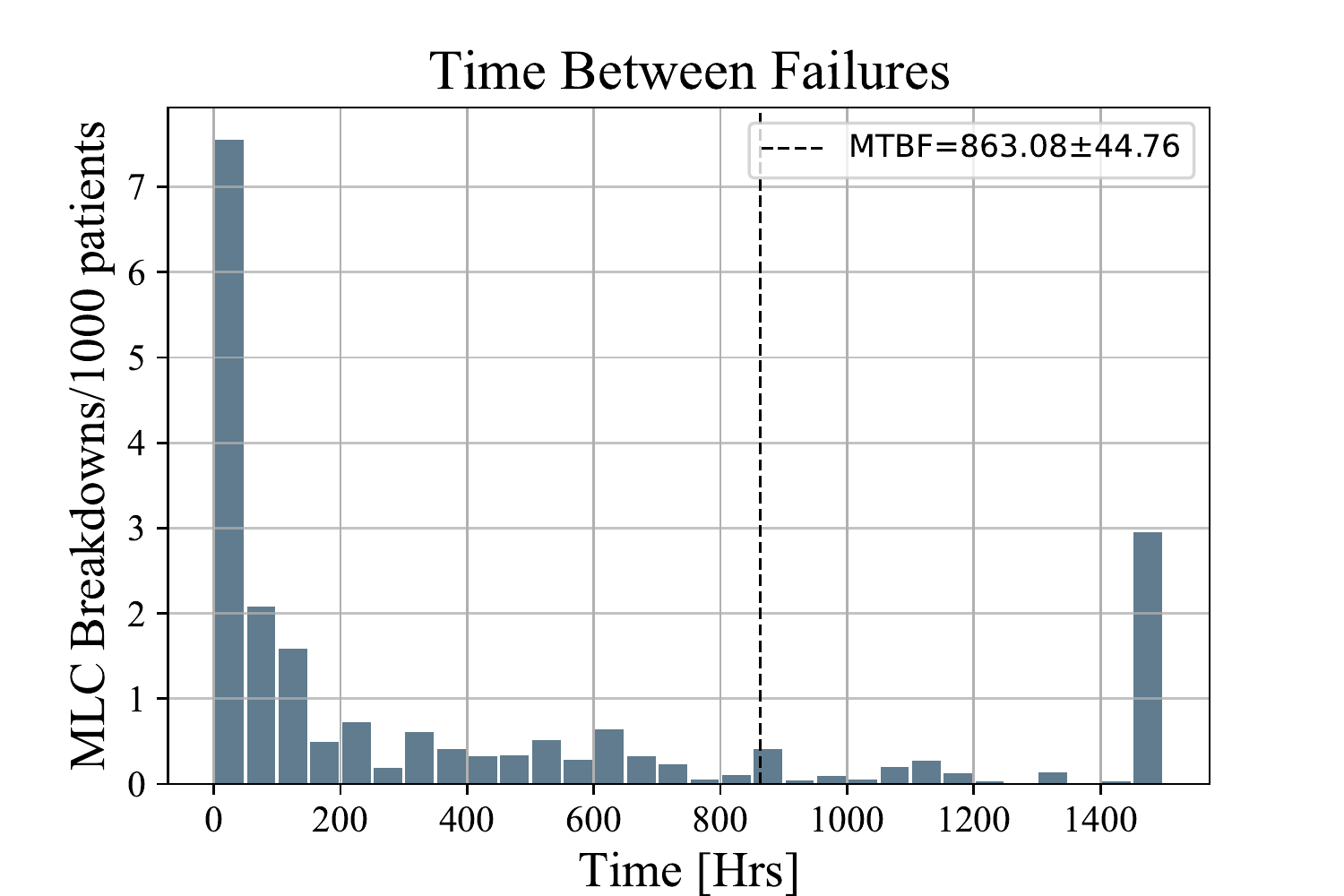}
	\end{subfigure}
	\caption{Histogram showing the time between failures for all LINACs (left) and just the MLC component (right). The Mean Time Between Failures (MTBF) is indicated by the dotted line. The overflow bin is 600 for the left and 1500 for the right.}
	\label{fig:MTBF}
\end{figure*}

\begin{figure*}[h]
		\centering
		\includegraphics[scale=0.45]{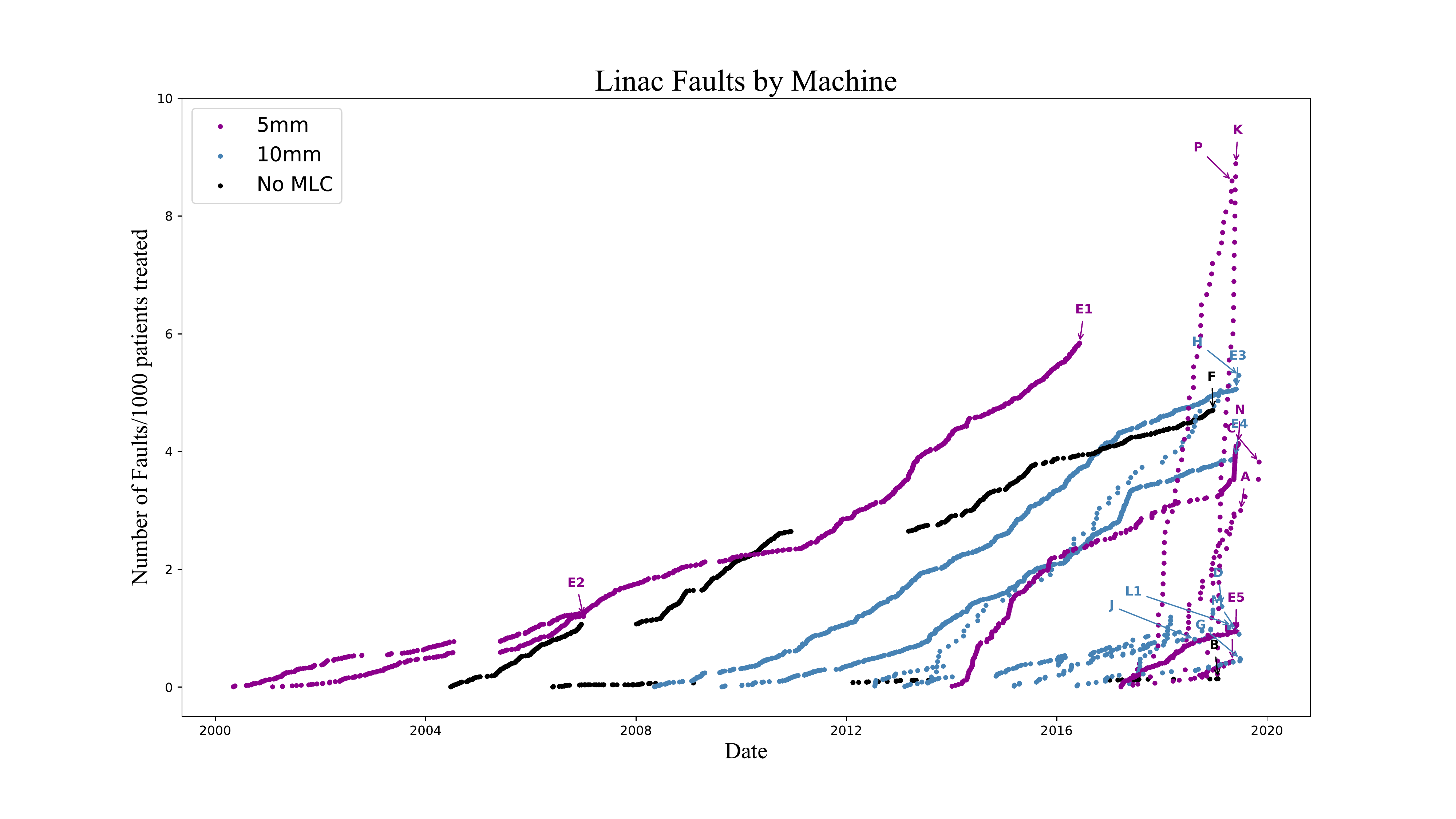}
		\caption{All faults displayed as a function of date per each hospital and grouped by the leaf width in the MLC.}
		\label{fig:AllFaults}
	\end{figure*}

LINAC performance is quantified by the downtime of all the machines and the Mean Time Between Failure (MTBF). MTBF is a parameter regularly used in engineering reliability analysis to quantify and study faults~\cite{MTBFSpeaks,Lienig}. The MTBF measures the average time between the resolution of one fault and the reporting of a subsequent one. The MTBF is defined as:
	\begin{align}
		MTBF = \int^{\infty}_0 R(t)dt = \int^{\infty}_0 tf(t)dt,
	\end{align}
where R(t) is called the reliability function and is re-expressed as the expected value of a probability density function of time to failure, f(t). This function only considers a machine's ``useful life"~\cite{MTBFSpeaks,EventHelix}, since early failures and wear out due to fatigue can affect the mean. From installation to decommissioning, one expects to see high failure rates early on and later on in a machine's lifetime and a lower constant failure rate through the middle. This is known as a 'Bathtub Curve'.\\ \\
From the data, the MTBF for a LINAC in Indonesia is 341.58 Hours, or about 14 days as seen in Fig \ref{fig:MTBF}. For an MLC in Indonesia, the MTBF is 863.08 Hours, or a little over a month (35 days). 
\\
Another way to understand this data is to look at the impact caused by these failures by dividing the total downtime of a machine by its total uptime. This can be seen in Table \ref{table:indomlcdata}. Normalising by the uptime per machine avoids misrepresenting the failures of one machine as a failure of other LINACs in the same hospital. Similarly, the contribution to the downtime due to MLC related faults can also be found in Table \ref{table:indomlcdata}. This is calculated by dividing the total downtime due to the MLC from the total downtime of the LINAC. \\
\begin{figure*}[h]
	\begin{subfigure}
  		\centering
    	\includegraphics[scale=0.55]{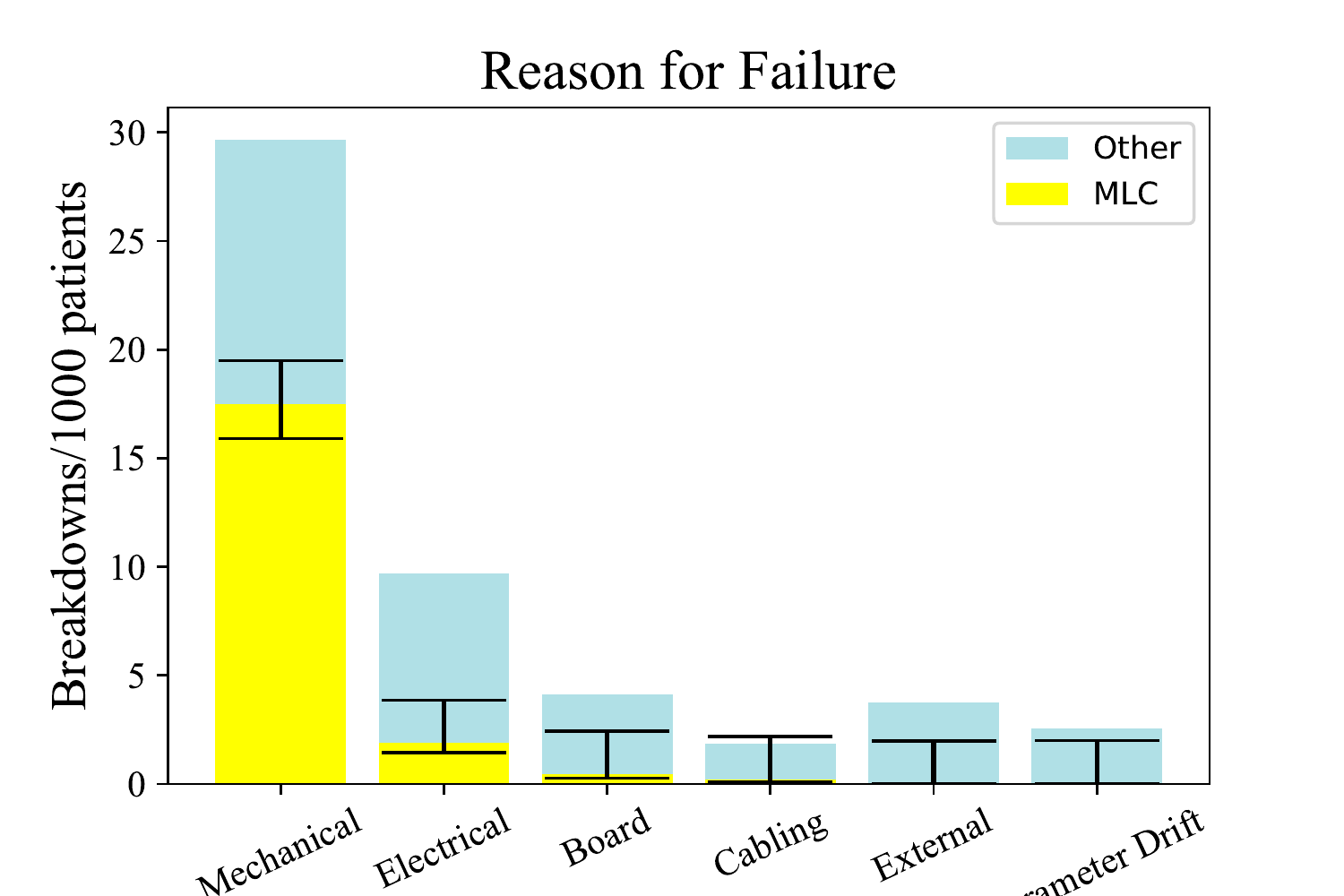}
	\end{subfigure}
	\begin{subfigure}
  		\centering
    	\includegraphics[scale=0.55]{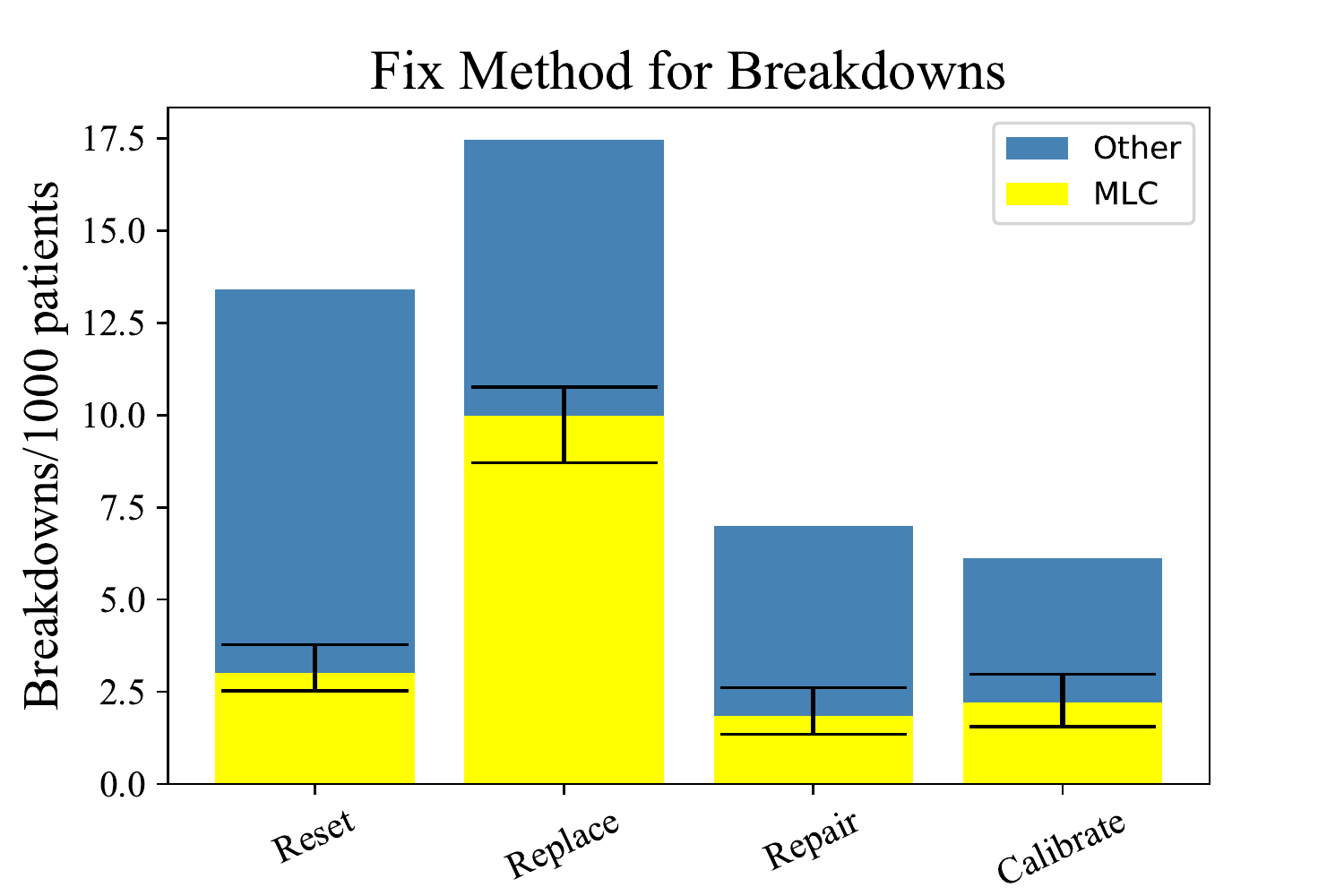}
	\end{subfigure}
	\caption{Histogram showing the most common reasons for failures (left) and the most common methods for fixing failures (right) across all LINACs in the Indonesian dataset normalised to the number of patients treated. The error bars are generated from the uncategorised data points.}
	\label{fig:ReasonFix}
\end{figure*}
\\
Next, we look at the cause of faults of LINAC as a whole and the MLC. As shown in Figure \ref{fig:ReasonFix}, mechanical faults are the most common type of faults with 29.66 mechanical faults occurring for every 1000 patients treated in Indonesia. Of these, ${59.02}_{-1.61}^{+1.98}\%$ are related to the MLC subsystem. \\ \\
Within the data, some faults and resolutions were left unfilled or lacked enough description to be categorised. We have used these data points as representative of errors in both recording and categorising faults, in order to produce the error bars in Figure \label{fig:ReasonFix}, in the following way. The maximum possible contribution to the results from these uncategorised data points would be if they all belonged to the MLC, this would add a quantity $\Delta_{MLC}$ to the number of MLC faults and is used to calculate the upper bound error bar, as per below. The minimum possible contribution would be if all uncategorised faults were in the LINAC, but not the MLC subsystem, which would add $\Delta_{LINAC}$ faults to the total system (but zero to the MLC). The minimum and maximum error bars are thus calculated as:	
\begin{align}
		min = \frac{F_{MLC}}{F_{LINAC}+\Delta_{LINAC}} && max = \frac{F_{MLC}+\Delta_{MLC}}{F_{LINAC}},
	\end{align}
where $F_{MLC}$ are the number of MLC faults of a given subcategory and $F_{LINAC}$ are all faults of a given subcategory. $\Delta_{LINAC}$ and $\Delta_{MLC}$ are the number of uncategorised faults of the whole dataset and MLC subset respectively. This is why the error bars of \emph{Cabling} in Fig \ref{fig:ReasonFix} exceed the height of the bar.\\ \\
The most common method of fixing a faulty LINAC is by replacing a component: this occurs 17.46 times for every 1000 patients treated. ${57.14}_{-1.27}^{+0.78} \%$ of the replacements are due to the MLC. The most common components being replaced in the MLC are leaves, leaf motors and T-Nuts, which are utilised in a mechanical context. The majority of replaced parts outside the MLC are fuses and cables.\\ \\ 
Figure \ref{fig:AllFaults} compiles all the data normalised to the number of patients treated and sorted by the width of the innermost leaves in the MLC. This gives the `Cumulative Bathtub Curve'. To compare the data appropriately, the number of faults have been normalised by the number of patients treated at each facility. Wroe's paper normalised their data by hours of uptime, though they stated normalising by patients treated is a better metric. This is true to a degree as hospitals which do not treat a large number of patients will become outliers, since the normalisation will heavily affect their faults per 1000 patients. This is clear in the two hospitals with near vertical lines (Hospital \textbf{K} and \textbf{P}). Hospital \textbf{C} also has a steep gradient due to the small number of patients, though it is harder to see in the figure.

%% file: sections/Discussion/discussion.tex
The data used in this analysis relies on accurate log keeping and consistent entries of failure of LINACs in a facility. All the conclusions drawn from the data have relied on thorough interactions with the hospital staff who recorded it at a dedicated workshop, but in most cases this level of data accuracy is not guaranteed, which is one reason why such a study has not previously been produced. Even so, data entries varied significantly in the level of detail provided and often had years worth of data missing as evidenced by the large gaps in Figure \ref{fig:AllFaults}. In Table \ref{table:indomlcdata}, for instance, Hospital \textbf{F}'s downtime was not possible to calculate since no rectification time was provided. Standardising or automating failure loggings would improve future studies. \\ \\
The faults over time shown in Figure \ref{fig:AllFaults} do not show the expected shape for a cumulative bathtub curve. This suggests either that machines are being used longer than their suggested lifetime~\cite{Vano} or that the failure rates in LMICs do not ease up over time. The reality may be that both factors are at play. It is well known in HIC's that due to advances in technology and software, LINACs get updates and replaced sooner than in LMICs. \\ \\
With improved or automated record-keeping, future studies could investigate are the mean times between specific failures. An initial analysis suggests a build-up of minor faults (Type \textbf{A} and \textbf{B} or Reset and Calibrate) before a disabling fault (Type \textbf{C} or Replace). To conclusively state this, however, would require a more rigorous record of minor faults. Anecdotally, the issue of comprehensive record-keeping and its value in improving hospital-led interventions for LINAC operation and preventative maintenance was raised on multiple occasions by the Indonesian participants in the workshop. \\ \\
A majority of the downtime, especially for replacements, is spent waiting either for spare parts or for Vendor engineers. In the survey, medical physicists at the various hospitals indicated that better training, better availability of vendor engineers and easier access to spare parts could improve uptime. \\ \\
From Figure \ref{fig:MTBF}, the MTBF for LINACs through Indonesia was calculated to be 341.59$\pm$12.19 Hrs. However, the data is a heavily skewed distribution and as such a better measure of the centre is the median which in this case is 52.5 Hrs or 2.19 days between failures.

\subsection{The Multi-Leaf Collimator}
The MLC contributes to an alarming portion of failures of LINACs in Indonesia. Of the machines which have an MLC, 25$\pm$6.37\% of the downtime and 27.3\% of the faults by number are due to the MLC. The downtime contribution being lower than the fault number contribution suggests MLCs take less time to repair than other faults. A majority of the MLC failures are mechanical in nature because, of all the subsystems in a LINAC, the MLC has the most moving parts, with at least 58 leaves, each with its own motor. To this effect, we can look at the contribution to downtime by the MLC from Table \ref{table:indomlcdata} as a function of leaf width. Omitting the outliers 83.9\% from \textbf{E4} and \textbf{K}), 5mm leaf widths account for 18.27$\pm$6.5\% (N = 8) of LINAC faults while 10mm leaves contribute 15.87$\pm$4.3\% (N = 7). Even though the average of the 10mm is lower, the spread of data is large due to low statistics. \\ \\
Comparing machines with and without MLCs to quantify differences in failure rates and downtimes is not possible in this dataset since only two machines don't have a MLC. One of which, Hospital \textbf{F}, has provided no downtimes. Instead, a comparison between Wroe's findings and these results can be investigated. The mean downtime for MLCs for data collected in the UK is 271.3 minutes compared to the 1159.34$\pm$338.19 minutes in Indonesia. This corresponds to an MLC downtime 4.3 times longer in the studied LMIC compared to the HIC. Another point of difference is that in the UK, type \textbf{B} and \textbf{C} MLC faults share around 45\% of the total downtime, whereas in Indonesia, it is dominated by type \textbf{C} faults at 96.6\%. Though this is in part due to longer wait times for spare parts and extended repair times, the high percentage is also due to a lack of type \textbf{A} \& \textbf{B} records.

\subsection{MLC Alternatives}
With the problems in the MLC persisting through all regions of the world~\cite{Atun,Pistenmaa,Wroe,Reichenvater,Balogun}, it is important to look at the possible alternatives for X-ray collimation. Prior to the introduction of Multi-leaf Collimators, LINACs used alloy block field shaping, however reverting to this method means an increase in treatment time and a reduction in prescribed dose being delivered~\cite{Galvin,Adams}. A similar method uses solid compensators to shape the beam, which is a simpler, yet non-automated delivery technique. The solid compensator when compared to the MLC results in fewer monitor units and a reduction in delivery time by half for complex treatment plans~\cite{Khadija,Kuros-Zolnierczuk}. For simple IMRT plans however, the MLC showed slightly better treatment plan quality. The caseload and complexity of treatments is thus important in considering the choice of collimation in future. One complaint of solid compensators is that they are labour intensive, however patient throughput is about the same and sometimes faster than MLC based treatments~\cite{Khadija}.\\ \\
An emerging alternative for X-ray collimation is the Scanning Pencil-beam High-speed Intensity- modulated X-ray source (SPHINX) collimator which uses a 10cm tungsten block with tumour size dependent tapered and diverging channels~\cite{Maxim2019,Trovati}. This removes the need for an MLC by driving the electrons to the most appropriate point on a bremsstrahlung target and having the diverging channels deliver the X-rays to the tumour- effectively shaping the radiation to the tumour further upstream. The applicability of this technology should be studied in terms of its use in more challenging maintenance environments in LMICs.\\ \\
An alternative that might cause the least amount of disruption to workflow is changing the parameters of the leaves in an MLC. Figure \ref{fig:ReasonFix} shows mechanical faults as the prevailing issue, and it is established that leaf motors needing to be replaced accounts for a majority of the replacement resolutions. Having fewer leaves would decrease the number of leaf motors used and decrease the total contact surface area between leaves in an MLC. How this will affect treatment planning quality is dependent on various factors such as treatment delivery technique and tumour volume~\cite{Yoganathan,Wu,Wang}.